\documentclass[twocolumn,twoside,slac]{revtex4}
\usepackage{graphicx}
\usepackage{fancyhdr}
\begin{document}

\title{A Measure of the Goodness of Fit in Unbinned Likelihood Fits; End of Bayesianism?}

%

\author{Rajendran Raja}
\affiliation{Fermilab, Batavia, IL 60510, USA}

\begin{abstract}
Maximum likelihood fits to data can be done using binned data
(histograms) and unbinned data. With binned data, one gets not only
the fitted parameters but also a measure of the goodness of fit. With
unbinned data, currently,  the fitted parameters are obtained but
no measure of  goodness of fit is available. This remains, to date, an
unsolved problem in statistics. Using Bayes' theorem and likelihood
ratios, we provide a method by which both the fitted quantities and a
measure of the goodness of fit are obtained for unbinned likelihood
fits, as well as errors in the fitted quantities. The quantity, 
conventionally interpreted as a Bayesian prior, is seen in this 
scheme to be a number not a distribution, that is determined from data.
\end{abstract}

\maketitle

\thispagestyle{fancy}
\section{Introduction}
As of the Durham conference~\cite{durham}, the problem of obtaining a
goodness of fit in unbinned likelihood fits was an unsolved one. In
what follows, we will denote by the vector $s$, the theoretical
parameters ($s$ for ``signal'') and the vector $c$, the experimentally
measured quantities or ``configurations''. For simplicity, we will
illustrate the method where both $s$ and $c$ are one dimensional,
though either or both can be multi-dimensional in practice. We thus
define the theoretical model by the conditional probability density
$P(c|s)$. Then an unbinned maximum likelihood fit to data is obtained
by maximizing the likelihood~\cite{fisher},
\begin{equation}
{\cal L} = \prod_{i=1}^{i=n} P(c_i|s)
\end{equation}
where the likelihood is evaluated at the $n$ observed data points
$c_i, i=1,n$.  Such a fit will determine the maximum likelihood value
$s^*$ of the theoretical parameters, but will not tell us how good the
fit is.  The value of the likelihood ${\cal L}$ at the maximum
likelihood point does not furnish a goodness of fit, since the
likelihood is not invariant under change of variable. This can be seen
by observing that one can transform the variable set $c$ to a variable
set $c'$ such that $P(c'|s^*)$ is uniformly distributed between 0 and
1. Such a transformation is known as a hypercube transformation, in
multi-dimensions. Other datasets will yield different values of
likelihood in the variable space $c$ when the likelihood is computed
with the original function $P(c|s^*)$. However, in the original
hypercube space, the value of the likelihood is unity regardless of
the dataset $c'_i,i=1,n$, thus the likelihood ${\cal L}$ cannot
furnish a goodness of fit by itself, since neither the likelihood, nor
ratios of likelihoods computed using the same distribution $P(c|s^*)$
is invariant under variable transformations. The fundamental reason
for this non-invariance is that only a single distribution, namely,
$P(c|s^*)$ is being used to compute the goodness of fit.

\section{Likelihood ratios}
In binned likelihood cases, where one is comparing a theoretical
distribution $P(c|s)$ with a binned histogram, there are two
distributions involved, the theoretical distribution and the data
distribution. The $pdf$ of the data is approximated by the bin
contents of the histogram normalized to unity. If the data consists of
$n$ events, the  $pdf$ of the data $P^{data}(c)$ is defined in the frequentist
sense as the normalized density distribution in $c$ space of $n$ events 
as $n\rightarrow \infty$. In the binned case, we can bin in finer and 
finer bins as $n\rightarrow \infty$ and obtain a smooth function, 
which we define as the $pdf$ of the data $P^{data}(c)$. In practice, one is always 
limited by statistics and the binned function will be an approximation 
to the true $pdf$. We can now define a likelihood ratio $\cal L_R$ such that
\begin{equation}
{\cal L_R} = \frac{\prod_{i=1}^{i=n} P(c_i|s)}{\prod_{i=1}^{i=n} P^{data}(c_i)}
\equiv \frac{P({\bf c_n}|s)}{P^{data}({\bf c_n})}
\end{equation}
where we have used the notation ${\bf c_n}$ to denote 
the event set $c_i, i=1,n$. Let us now note that ${\cal L_R}$ is invariant 
under the variable transformation $c\rightarrow c'$, since 
\begin{eqnarray}
 P(c'|s) = |\frac{dc}{dc'}|P(c|s) \\
 P^{data}(c') = |\frac{dc}{dc'}|P^{data}(c) \\
 {\cal L'_R} = {\cal L_R}
\end{eqnarray}
and the Jacobian of the transformation $|\frac{dc}{dc'}|$
cancels in the numerator and
denominator in the ratio. This is an extremely important property of
the likelihood ratio ${\cal L_R}$ that qualifies it to be a goodness
of fit variable. Since the denominator  $P^{data}({\bf c_n})$ is 
independent of the theoretical parameters $s$, both the likelihood 
ratio and the likelihood maximize at the same point $s^*$. One can also 
show~\cite{raja1} that the maximum value of the likelihood ratio occurs 
when the theoretical likelihood $P(c_i|s)$ and the data 
likelihood $P^{data}(c_i)$ are equal for all $c_i$.
\section{Binned Goodness of Fit}
In the case where the $pdf$ $P^{data}(c)$ is estimated by binned
histograms and the statistics are Gaussian, it is readily
shown~\cite{raja1} that the commonly used goodness of fit variable
 $\chi^2= -2 log {\cal L_R}$.
 It is worth emphasizing that the likelihood
ratio as defined above is needed and not just the negative log of
theoretical likelihood $P({\bf c_n}|s)$ to derive this result. The popular 
conception that $\chi^2$ is -2 log $P({\bf c_n}|s)$ is simply incorrect!.
It can also be shown that the likelihood ratio defined above can describe 
the binned cases where the statistics are Poissonian~\cite{raja2}. In order 
to solve our problem of goodness of fit in unbinned likelihood cases, one 
needs to arrive at a method of estimating the data $pdf$ $P^{data}(c)$
without the use of bins.
\section{Unbinned Goodness of Fit}
One of the better known methods of estimating the probability density of 
a distribution in an unbinned case is by the use of Probability 
Density Estimators $(PDE's)$, also known as Kernel Density Estimators~\cite{parzen} $(KDE's)$. The $pdf$ $P^{data}(c)$ is approximated by
\begin{equation}
  P^{data}(c)\approx PDE(c) = \frac{1}{n}\sum_{i=1}^{i=n} {\cal G}(c-c_i)
\label{pde}
\end{equation}
where a Kernel function ${\cal G}(c-c_i)$ is centered around each data 
point $c_i$, is so defined that it normalizes to unity and for 
large $n$ approaches a Dirac delta function~\cite{raja1}. The choice of the 
Kernel function can vary depending on the problem. A popular kernel is the 
Gaussian defined in the multi-dimensional case as
\begin{equation}
 {\cal G}(c) = \frac{1}{(\sqrt{2\pi}h)^d\sqrt(det(E))}
exp(\frac{-H^{\alpha\beta}c^\alpha c^\beta}{2h^2})
\end{equation}
where $E$ is the error matrix of the data defined as
\begin{equation}
 E^{\alpha,\beta} = <c^\alpha c^\beta > -<c^\alpha><c^\beta>
\end{equation}
and the $<>$ implies average over the $n$ events, and $d$ is the
 number of dimensions. The Hessian matrix $H$ 
is defined as the inverse of $E$ and 
the repeated indices imply summing over.  The parameter $h$ is a 
``smoothing parameter'',  which has\cite{hoptim} a 
suggested optimal value $h \propto n^{-1/(d+4)}$, that 
satisfies the asymptotic condition 
\begin{equation}
  {\cal G}_\infty(c-c_i)\equiv\lim_{n \rightarrow \infty} {\cal G}(c-c_i) = \delta (c-c_i)
\end{equation}
The parameter $h$ will depend on the local number density and will have to be 
adjusted as a function of the local density to obtain good representation of 
the data by the $PDE$.
Our proposal for the goodness of fit in unbinned likelihood fits is thus 
the likelihood ratio
\begin{equation}
{\cal L_R} = \frac{P({\bf c_n}|s)}{P^{data}({\bf c_n})} \approx 
\frac{P({\bf c_n}|s)}{P^{PDE}({\bf c_n})}
\end{equation}
evaluated at the maximum likelihood point $s^*$.
\section{An illustrative example}
\begin{figure}[t]
\centerline{\includegraphics[width=65mm]{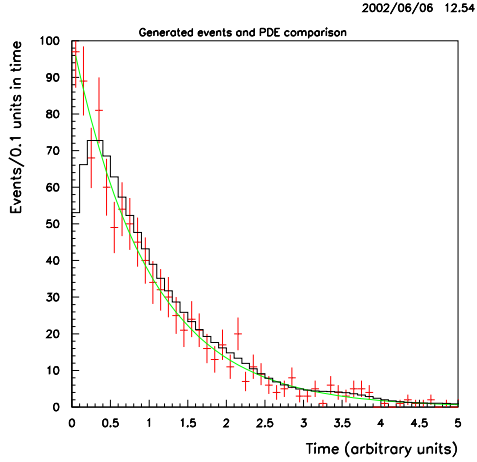}}
\caption
{Figure shows the histogram (with errors) of generated events. 
Superimposed is the theoretical
curve $P(c|s)$  and the $PDE$ estimator (solid) histogram with no errors.
\label{genev}}
\end{figure}

We consider a simple one-dimensional case where the data is an exponential distribution, say decay times of a radioactive isotope. 
The theoretical prediction is given by
\begin{equation}
P(c|s) = \frac{1}{s}\exp(-\frac{c}{s})
\end{equation}
We have chosen an exponential with $s=1.0$ for this example.
The Gaussian Kernel for the $PDE$ would be given by
\begin{equation}
{\cal G}(c) = \frac{1}{(\sqrt{2\pi} \sigma h)}
\exp (-\frac{c^2}{2\sigma^2h^2})
\end{equation}
where the variance $\sigma$ of the exponential is numerically equal to $s$.
To begin with, we chose a constant value for the smoothing parameter, 
which for 1000 events generated is calculated to be 0.125.
Figure~\ref{genev} shows the generated events, the theoretical curve
$P(c|s)$ and the $PDE$ curve $P(c)$ normalized to the number of
events. The $PDE$ fails to reproduce the data near the origin due to the 
boundary effect, whereby the 
Gaussian probabilities for events close to the origin 
spill over to negative values of $c$. 
This lost probability would be compensated by events on the
exponential distribution with negative $c$ if they existed. In our
case, this presents a drawback for the $PDE$ method, which we will
remedy later in the paper using $PDE$ definitions on the  hypercube
and periodic boundary conditions. For the time being, we will confine our 
example to values of $c > 1.0$ to avoid the boundary effect. 

\begin{figure}[tbh!]
\centerline{\includegraphics[width=65mm]{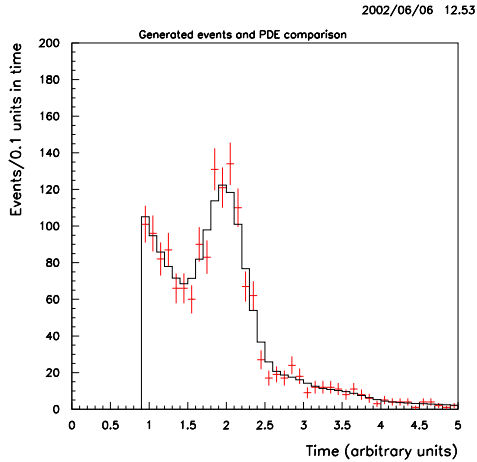}}
\caption[generated events]
{Figure shows the histogram (with errors) of 1000 events in the 
fiducial interval $1.0<c<5.0$ generated as
an exponential with decay constant $s$=1.0 with a superimposed
Gaussian of 500 events centered at $c$=2.0 and width=0.2.  
The $PDE$ estimator is
the (solid) histogram with no errors. 
\label{genev1}}
\end{figure}

 In order to test the goodness of fit 
capabilities of the likelihood ratio ${\cal L_R}$, we superimpose a 
Gaussian on the exponential and try and fit the data by a simple exponential.
Figure~\ref{genev1} shows the ``data'' with 1000 events generated as
an exponential in the fiducial range $1.0 <c< 5.0$. Superimposed on it 
is a Gaussian of 500 events. More events in the exponential are generated 
in the interval $0.0 <c < 1.0$ to avoid the boundary effect at the 
fiducial boundary at c=1.0.
Since the number
density varies significantly, we have had to introduce a method of
iteratively determining the smoothing factor as a function of $c$ as
described in~\cite{raja1}. With this modification in the $PDE$, one
gets a good description of the behavior of the data by the $PDE$ as
shown in Figure~\ref{genev1}.
We now vary the number of events in the Gaussian and obtain the value
of the negative log likelihood ratio ${\cal NLLR}$ as a function of
the strength of the Gaussian.  Table~\ref{tab1} summarizes the
results. The number of standard deviations the unbinned likelihood fit is from
what is expected is determined empirically by plotting the value of
${\cal NLLR}$ for a large number of fits where no Gaussian is
superimposed (i.e. the null hypothesis) and determining the mean and
$RMS$ of this distribution and using these to estimate the number of
$\sigma$'s the observed ${\cal NLLR}$ is from the null
case. Table~\ref{tab1} also gives the results of a binned fit on the
same ``data''. It can be seen that the unbinned fit gives a $3\sigma$
discrimination when the number of Gaussian events is 85, where as the
binned fit gives a $\chi^2/ndf$ of 42/39 for the same case. We intend to
 make these tests more sophisticated in future work.
\begin{table}[bht!]
\caption{
Goodness of fit results from unbinned likelihood and binned likelihood fits for
various data samples. The negative values for the number of 
standard deviations in some of the examples is due to statistical fluctuation.
\label{tab1}}
\centering\leavevmode
\begin{tabular}{|c|c|c|c|}
\hline
Number of & Unbinned fit & Unbinned fit& Binned fit $\chi^2$ \\
Gaussian events & ${\cal NLLR}$ & $N\sigma$ & 39 d.o.f.\\
\hline
500 & 189. & 103 & 304 \\
250 & 58.6 & 31 & 125 \\
100 & 11.6 & 4.9& 48 \\
85 & 8.2 & 3.0 & 42 \\
75 & 6.3 & 1.9 & 38 \\
50 & 2.55 & -0.14 &30 \\
0  & 0.44 & -1.33 & 24 \\
\hline
\end{tabular}
\end{table}

\begin{figure}[tbh!]
\centerline{\includegraphics[width=65mm]{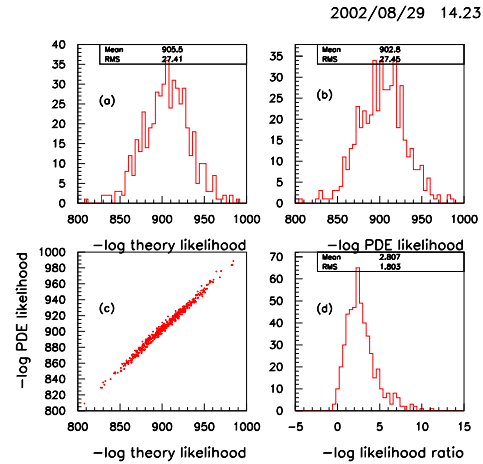}}
\caption
{(a) shows the distribution of the negative log-likelihood -$log_e
(P({\bf c_n}|s))$ for an ensemble of experiments where data and experiment 
are expected to fit. (b) Shows the negative log $PDE$ likelihood -$log_e
(P({\bf c_n}))$ for the same data (c) Shows the correlation between the
two and (d) Shows the negative log-likelihood ratio ${\cal NLLR}$ that
is obtained by subtracting (b) from (a) on an event by event basis.
\label{compar}}
\end{figure}

 Figure~\ref{compar} shows the variation of -log $P({\bf c_n}|s)$ and
-log $P^{PDE}({\bf c_n})$ for an ensemble of 500 experiments each with
the number of events $n=1000$ in the exponential and no events in the
Gaussian (null hypothesis). It can be seen that -log $P({\bf c_n}|s)$ and
-log $P^{PDE}({\bf c_n})$ are correlated with each other and the difference 
between the two (-log ${\cal NLLR}$) is a much narrower distribution 
than either and provides the goodness of fit discrimination.
\subsection{Improving the $PDE$}
The $PDE$ technique we have used so far suffers from two drawbacks;
firstly, the smoothing parameter has to be iteratively adjusted
significantly over the full range of the variable $c$, since the
distribution $P(c|s)$ changes significantly over that range; and
secondly, there are boundary effects at $c$=0 as shown in
figure~\ref{genev}.  Both these flaws are remedied if we define the
$PDE$ in hypercube space.  After we find the maximum likelihood point
$s^*$, for which the $PDE$ is not needed, we transform the variable
$c\rightarrow c'$, such that the distribution $P(c'|s^*)$ is flat and
$0<c'<1$. The hypercube transformation can be made even if $c$ is 
multi-dimensional by initially going to a set of variables that are
 uncorrelated and then making the hypercube transformation. 
The transformation can be such that any interval in $c$ space maps on
to the interval $(0,1)$ in hypercube space.  We solve the boundary
problem by imposing periodicity in the hypercube.  In the one
dimensional case, we imagine three ``hypercubes'', each identical to the
other on the real axis in the intervals $(-1,0)$, $(0,1)$ and $(1,2)$. The
hypercube of interest is the one in the interval $(0,1)$.  When the
probability from an event kernel leaks outside the boundary $(0,1)$, we
continue the kernel to the next hypercube. Since the hypercubes are
identical, this implies the kernel re-appearing in the middle
hypercube but from the opposite boundary. Put mathematically, the kernel 
is defined such that
\begin{eqnarray}
 {\cal G}(c'-c'_i) ={\cal G}(c'-c'_i-1);\: c'>1 \\
 {\cal G}(c'-c'_i) ={\cal G}(c'-c'_i+1);\: c'<0 
\end{eqnarray}

Although a Gaussian Kernel will work on the hypercube, 
the natural kernel to use considering the shape of the hypercube  
would be  the function ${\cal G}(c')$
\begin{eqnarray}
 {\cal G}(c') = \frac{1}{h} ;\: |c'|<\frac{h}{2} \\
 {\cal G}(c') = 0 ;\: |c'|>\frac{h}{2} 
\end{eqnarray} 
This kernel would be subject to the periodic boundary conditions given above, 
which further ensure that every event in hypercube space is treated
 exactly as every other event irrespective of their co-ordinates.
The parameter $h$ is a smoothing parameter which needs to be chosen 
with some care. However, since the theory distribution is flat in 
hypercube space, the smoothing parameter may not need to be iteratively 
determined over hypercube space to the extent that 
data distribution is similar to 
the theory distribution. Even if iteration is used, the variation in $h$ 
in hypercube space is likely to be much smaller.
\begin{figure}[tbh!]
\centerline{\includegraphics[width=65mm]{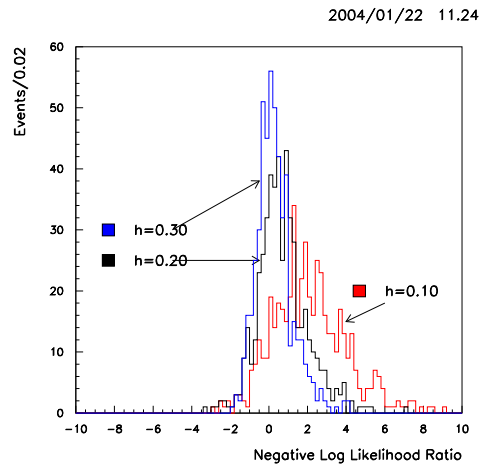}}
\caption
{The distribution of the negative log likelihood ratio ${\cal NLLR}$ for 
the null hypothesis for an ensemble of 500 experiments each with 1000 events, 
as a function of the smoothing factor  $h$=0.1, 0.2 and 0.3
\label{hyperbin}}
\end{figure}

\begin{figure}[tbh!]
\centerline{\includegraphics[width=65mm]{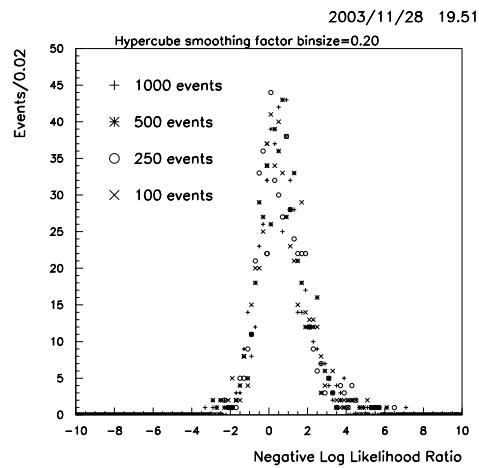}}
\caption
{The distribution of the negative log likelihood ratio ${\cal NLLR}$ for 
the null hypothesis for an ensemble of 500 experiments each with
the smoothing factor  $h$=0.2, as a function of the number of events 
\label{hyperev}}
\end{figure}

 Figure~\ref{hyperbin} shows the distribution of the ${\cal NLLR}$ for the 
null hypothesis for an ensemble of 500 experiments each with 1000 events as a 
function of the smoothing factor $h$. It can be seen that the distribution 
narrows considerably as the smoothing factor increases. We choose an operating
 value of 0.2 for $h$ and study the dependence of the ${\cal NLLR}$ as a 
function of the number of events ranging from 100 to 1000 events, as shown in 
figure~\ref{hyperev}. The 
dependence on the number of events is seen to be weak, indicating 
good behavior. The $PDE$ thus arrived computed with $h$=0.2 can be transformed
 from the hypercube space to $c$ space and will reproduce data smoothly and 
with no edge effects. We note that it is also easier to arrive at an 
analytic theory of ${\cal NLLR}$ with the choice of this simple kernel.
\section{End of Bayesianism?}
By Bayesianism, we mean the practice of ``guessing'' a prior distribution 
and introducing it into the calculations. In what follows we will show that 
what is conventionally thought of as a Bayesian prior distribution is 
in reality a number that can be calculated from the data. We are able to 
do this since we use two $pdf$'s, one for theory and one for data. In what follows, we
will interpret the probability distribution of the parameter $s$ in a 
strictly frequentist sense. The $pdf$ of $s$ is the distribution of 
the best estimator of the true value $s_T$ of $s$ from 
an ensemble of an infinite number of
identical experiments with the same statistical power $n$. 

\subsection{Calculation of fitted errors}
After the fitting is done and the goodness of fit is evaluated, one
needs to work out the errors on the fitted quantities. One needs to
calculate the posterior density $P(s|{\bf c_n})$, 
which carries information not 
only about the maximum likelihood point $s^*$, from a single experiment,
 but how  such a measurement is likely to fluctuate if we 
repeat the experiment. 
The joint probability density $P(s,{\bf c_n})$ of observing the
parameter  $s$ and the data ${\bf c_n}$ is given by

\begin{equation}
 P^{data}(s,{\bf c_n}) = P(s|{\bf c_n})P^{data}({\bf c_n})
\label{joint}
\end{equation}
where we use the superscript $^{data}$ to distinguish the joint probability
 $P^{data}(s,{\bf c_n})$ as having come from using the data $pdf$. 
If we now integrate the above equation over all possible datasets ${\bf c_n}$,
 we get the expression for the $pdf$ of $s$.
\begin{equation}
{\cal P}_n(s) = \int P^{data}(s,{\bf c_n}) d{\bf c_n} = \int P(s|{\bf c_n})P^{data}({\bf c_n}) d{\bf c_n}
\label{pns}
\end{equation}
where we have used the symbol ${\cal P}$ to distinguish the fact that
it is the true $pdf$ of $s$ obtained from an infinite ensemble.  We
use the subscript $n$ in ${\cal P}_n(s)$ to denote that the $pdf$ is
obtained from an ensemble of experiments with $n$ events each.  Later
on we will show that ${\cal P}_n(s)$ is indeed dependent on $n$.
Equation~\ref{pns} states that in order to obtain the $pdf$ of the
parameter $s$, one needs to add together the conditional probabilities
$P(s|{\bf c_n})$ over an ensemble of events, each such
distribution weighted by the ``data likelihood'' $P^{data}({\bf
c_n})$.  At this stage of the discussion, the functions
$P^{data}(s|{\bf c_n})$ are unknown functions. We have however worked
out ${\cal L_R}(s)$ as a function of $s$ and have evaluated the
maximum likelihood value $s^*$ of s.  We can choose an arbitrary value
of $s$ and evaluate the goodness of fit at that value using the
likelihood ratio. When we choose an arbitrary value of $s$, we are in
fact hypothesizing that the true value $s_T$ is at this value of
$s$. $L_R(s)$ then gives us a way of evaluating the relative goodness
of fit of the hypothesis as we change $s$. Let us now take an arbitrary
value of $s$ and hypothesize that that is the true value. Then the
joint probability of observing ${\bf c_n}$ and $s_T$ being at this
value of $s$ is given from the data end by equation~\ref{joint}.

Similarly, from the theoretical end, one can calculate the joint
probability of observing the dataset ${\bf c_n}$, with the true value
being at $s$.  The true value $s_T$ is taken to be the maximum
likelihood point of the $pdf$ ${\cal P}_n(s)$. It may coincide with
the mean value of the $pdf$ ${\cal P}_n(s)$.  These statements are
assertions of the unbiased nature of the data from the experiment.  At
this point, there is no information available on where the true value
$s_T$ lies. One can make the hypothesis that a particular value of $s$
is the true value and the probability of obtaining a best estimator $s^*$ from 
experiments of the type being performed in the interval 
$s_T$ and $s_T+ds_T$ is 
${\cal P}_n(s_T)ds_T$. 
The actual value of this number is a function of the
experimental resolution and the statistics $n$ of the experiment.
The  joint probability $P^{theory}(s,{\bf c_n})$ from the theoretical end
 is given by the product of the 
probability density of the $pdf$ of $s$ at the true value of $s$, 
namely ${\cal P}_n(s_T)$, and the theoretical likelihood $P(c_n|s)$
 evaluated at the true value, which by our hypothesis is $s$.
\begin{equation}
 P^{theor}(s,{\bf c_n}) = P^{theor}({\bf c_n}|s){\cal P}_n(s_T)
\end{equation}
The joint probability $P(s,{\bf c_n})$ is a joint distribution of 
the theoretical parameter $s$ and data ${\bf c_n}$. The two ways of evaluating 
this (from the theoretical end and the data end) must yield the 
same result, for consistency.
 This is equivalent to equating 
$ P^{data}(s,{\bf c_n})$ and $ P^{theor}(s,{\bf c_n})$. This gives the equation
\begin{equation}
  P(s|{\bf c_n})P^{data}({\bf c_n}) =  P^{theor}({\bf c_n}|s){\cal P}_n(s_T)
\label{bay2}
\end{equation}
which is a form of Bayes' theorem, but with two $pdf's$ (theory and data). 
Let us note that the above equation can be immediately re-written as 
a likelihood ratio
\begin{equation}
 {\cal L_R} = \frac{P(s|{\bf c_n})}{{\cal P}_n(s_T)} = 
 \frac{ P^{theor}({\bf c_n}|s)}{P^{data}({\bf c_n})}
\end{equation}
which is what is used to obtain the goodness of fit. 
In order to get 
the fitted errors, we need to evaluate $ P(s|{\bf c_n})$ which necessitates a
 better understanding of what ${\cal P}_n(s_T)$ is in equation~\ref{bay2}. Rearranging 
equation~\ref{bay2}, one gets
\begin{equation}
  P(s|{\bf c_n}) = {\cal L_R}(s){\cal P}_n(s_T) =  
\frac{ P^{theor}({\bf c_n}|s)}{P^{data}({\bf c_n})}{\cal P}_n(s_T)
\label{lreq}
\end{equation}
\subsubsection{To show that ${\cal P}_n(s_T)$ depends on n}
In practice, in both the binned and unbinned cases, one only has an
approximation to $P^{data}({\bf c_n})$.  As $n \rightarrow \infty$, in
the absence of experimental bias, one expects to determine the
parameter set $s$ to infinite accuracy; and $ P(s|{\bf
c_n})\rightarrow \delta(s-s_T)$, where $s_T$ is the true value of $s$.
However, for the null hypothesis, as $n\rightarrow \infty$, the
statistical error introduced by our use of $PDE$ in the unbinned case
or by binning in the binned case becomes negligible with the result that 
the theory $pdf$ describes the data for all $c$ at the true value $s_T$. i.e.
\begin{equation}
\frac{ P^{theor}(c|s_T)}{P^{data}(c)} 
\rightarrow 1\: as\:n\rightarrow \infty
\end{equation}
When one evaluates the likelihood ratio ${\cal L_R}$ over $n$ events, 
with $n\rightarrow \infty$, the likelihood ratio does not necessarily remain 
unity. This is due to fluctuations in the data which grow as $\sqrt(n)$. 
For the binned likelihood case with  $n_b$ bins, one can show that as 
$n\rightarrow \infty$,
\begin{equation}
{\cal L_R}\rightarrow e^{-\sum_{i=1}^{i=n_b}\chi_i^2/2} \rightarrow e^{-n_b/2}
\end{equation}
This is just an example of the likelihood ratio theorem.  If one uses
a binned $\chi^2$ fit, which can also be thought of as maximizing a
likelihood ratio, one gets the same limit as when using binned
likelihood fits. The point is that ${\cal L_R}$ is finite as
$n\rightarrow \infty$.  In the unbinned case, we have currently no
analytic theory available. However, one can argue that the binned case
with the number of bins $n_b\rightarrow
\infty$ and $n_b<<n$ should approach the unbinned limit.  In this
case, the unbinned ${\cal L_R}$ also is finite for infinite
statistics.  This implies that ${\cal P}_n(s_T)\rightarrow \infty$ as
$n\rightarrow \infty$.  i.e ${\cal P}_n(s_T)$ depends on $n$.
This puts an end to the notion of a monolithic
Bayesian prior interpretation for ${\cal P}_n(s)$.
\subsubsection{To show that ${\cal P}_n(s_T)$ is constant with respect to $s$}
\begin{figure}[tbh]
\centerline{\includegraphics[width=65mm]{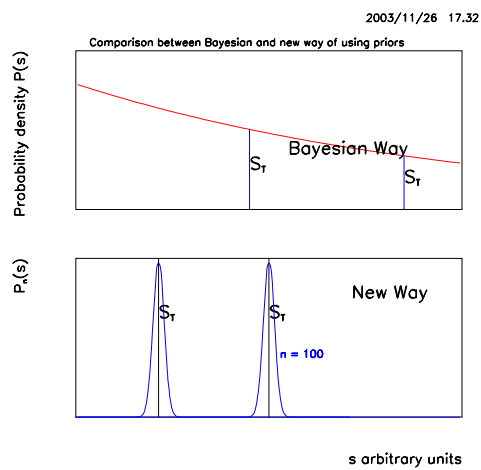}}
\caption[]
{Comparison of the usage of Bayesian priors with the new method. In the upper
figure, illustrating the Bayesian method, 
an unknown distribution is guessed at by the user based on
``degrees of belief'' and the value of the Bayesian prior changes as
the variable $s$ changes. In the lower figure, an ``unknown
concomitant'' distribution is used whose shape depends on the
statistics. In the case of no bias, this distribution peaks at the
true value of $s$. As we change $s$, we change our hypothesis as to 
where the true value of $s$ lies, and the distribution shifts with $s$
as explained in the text. The value of the distribution at the true
value is thus independent of $s$.}

\label{bnb}
\end{figure}

  When one varies the likelihood ratio in
equation~\ref{lreq} as a function of $s$, for each value of $s$, one
is making a hypothesis that $s=s_T$. As one changes s, a new
hypothesis is being tested that is mutually exclusive from the
previous one, since the true value can only be at one location. So as
one changes $s$, one is free to move the $distribution$ 
${\cal P}_n(s)$ so that $s_T$ is at the value of $s$ being
tested. This implies that ${\cal P}_n(s_T)$ does not change as one changes $s$
and is a constant $wrt$ s, which we can now write as
$\alpha_n$. Figure~\ref{bnb} illustrates these points graphically.
Thus ${\cal P}_n(s_T)$ in our equations is a number, not a function. 
The distribution ${\cal P}_n(s)$ should not be thought of 
 as a ``prior'' but as an ``unknown concomitant'', which
depends on the statistics and the measurement capabilities of the
apparatus. For a given apparatus, there are a denumerable infinity of such 
distributions, one for each $n$. These distributions become narrower as $n$ 
increases and ${\cal P}_n(s_T)\rightarrow \infty$ as $n\rightarrow \infty$. 

\subsection{New form of equations}
Equation~\ref{lreq} can now be re-written
\begin{equation}
  P(s|{\bf c_n}) = \frac{P({\bf c_n}|s)\alpha_n}{P^{data}({\bf c_n})}
\label{postn1}
\end{equation}
Since $P(s|{\bf c_n})$ must normalize to unity, one gets for $\alpha_n$,
\begin{equation}
\alpha_n = \frac{P^{data}({\bf c_n})}{\int P({\bf c_n}|s) ds} = 
\frac{1}{\int {\cal L_R}(s) ~ds}
\end{equation}
We have thus determined $\alpha_n$, the value of the ``unknown
concomitant'' at the true value $s_T$ using our data set $c_n$. This
is our $measurement$ of $\alpha_n$ and different datasets will give
different values of $\alpha_n$, in other words $\alpha_n$ will have a
sampling distribution with an expected value and standard
deviation. As $n\rightarrow \infty$, the likelihood ratio ${\cal L_R}$ 
will tend to a finite value at the true value and zero for all other values, 
and $\alpha_n \rightarrow \infty$ as a result.

Note that it is only possible to write down an expression
 for $\alpha_n$ dimensionally when a likelihood ratio ${\cal L_R}$ 
is available. 
This leads to
\begin{equation}
 P(s|{\bf c_n}) = \frac{{\cal L_R}}{\int {\cal L_R}~ds} = 
\frac {P({\bf c_n}|s)}{\int P({\bf c_n}|s)ds}
\label{poster}
\end{equation}
The last equality in equation~\ref{poster} is the same expression that
``frequentists'' use for calculating their errors after fitting,
namely the likelihood curve normalized to unity gives the parameter
errors. If the likelihood curve is Gaussian shaped, then this
justifies a change of negative log-likelihood of $\frac{1}{2}$ from
the optimum point to get the $1 \sigma$ errors. Even if it is not
Gaussian, as we show in section (\ref{secpost}),
we may use the expression for $P(s|{\bf c_n})$ as a $pdf$ of
the parameter $s$ to evaluate the errors.

%
%
The normalization condition
\begin{equation}
 P({\bf c_n}) = \int P^{theory}(s,{\bf c_n}) ds = 
\int P(c_n|s){\cal P}_n(s_T) ds
\end{equation}
is obeyed by our solution, since
\begin{equation}
 \int P({\bf c_n}|s){\cal P}_n(s_T)~ds = 
\int \alpha_n P({\bf c_n}|s)~ds \equiv P^{data}({\bf c_n})
\label{ident}
\end{equation}
The expression $\int \alpha_n P({\bf c_n}|s)~ds$ in the above equation
may be thought of as being due to an ``unknown concomitant'' whose
peak value is distributed uniformly in $s$ space. The likelihoods of
the theoretical prediction $P({\bf c_n}|s)$ contribute with equal
probability each with a weight $\alpha_n$, to sum up to form the data
likelihood $P^{data}({\bf c_n})$. i.e. the data, due to its statistical
inaccuracy will entertain a range of theoretical parameters. However,
equation~\ref{ident} does not give us any further information, since it is
obeyed identically.
Fitting for the maximum likelihood value $s^*$ of $s$ is attained by
maximizing the likelihood ratio ${\cal L_R} = \frac{P({\bf
c_n}|s)}{P^{data}({\bf c_n})}$. 
The goodness of fit is obtained using the value of ${\cal L_R}$ at
the maximum likelihood point. The best theoretical prediction is
$P(c|s^*)$, and this prediction is used to compare to the data $pdf$
$P^{data}(c)$. Note that the maximum likelihood value $s$ is also the
same point at which the posterior density $P(s|c)$ peaks. This is true
only in our method. When an arbitrary Bayesian prior is used, the
maximum likelihood value is not the same point at which the posterior
density will peak.
Note also that the normalization equation $\int {\cal P}_n(s)~ds$=1 is still
valid. The integral
\begin{equation}
 \int \alpha_n~ds \neq 1
\end{equation}
since $\alpha_n$ is our measurement of the value of ${\cal P}_n(s)$ at the
true value. It is a measure of the statistcal accuracy of the
experiment. The larger the value of $\alpha_n$, the narrower the
distribution ${\cal P}_n(s)$ and the more accurate the experiment.
\section{Combining Results of Experiments}
Each experiment should publish a likelihood curve for its fit as well as a 
number for the data likelihood $P^{data}({\bf c_n})$. Combining the results of 
two experiments with $m$ and $n$ experiments each, involves multiplying the 
likelihood ratios.
\begin{equation}
  {\cal L_R}_{m+n}(s) ={\cal L_R}_m(s) \times {\cal L_R}_n(s) = 
\frac{P({\bf c_m}|s)}{P^{data}({\bf c_m})} 
\times \frac{P({\bf c_n}|s)}{P^{data}({\bf c_n})}
\end{equation}
Posterior densities and goodness of fit can be deduced from the combined 
likelihood ratio.
\section{Interpreting the results of one experiment}
\label{secpost}
After performing a single experiment with $n$ events, we now can
calculate $P(s|{\bf c_n})$, using equation~\ref{poster}.
Equation~\ref{pns} gives the prescription for arriving at ${\cal
P}_n(s)$, given an ensemble of such experiments, the contribution from
each experiment being weighted by the ``data likelihood''
$P^{data}({\bf c_n})$ for that experiment. The ``data likelihoods''
integrate to unity, i.e $\int P^{data}({\bf c_n}) d{\bf c_n}$ = 1.  In
the case of only a single experiment, with the observed ${\bf c_n}$
being denoted by ${\bf c_n^{obs}}$,

\begin{equation}
 P^{data}({\bf c_n}) = \delta ({\bf c_n - c_n}^{obs})
\end{equation}
Equation~\ref{pns}, for a single experiment, then reduces to
\begin{equation}
{\cal P}_n(s) =  \int P(s|{\bf c_n})P^{data}({\bf c_n}) d{\bf c_n} =  P(s|{\bf c_n}^{obs})
\end{equation}
i.e. given a single experiment, the best estimator for ${\cal P}_n(s)$, 
the $pdf$ of $s$, is $P(s|{\bf c_n}^{obs})$ and thus the best
estimator for the true value $s_T$ is $s^{*obs}$ deduced from the
experiment. We can thus use $P(s|{\bf c_n}^{obs})$ as though it is the $pdf$
of $s$ and deduce limits and errors from it. The proviso is of course
that these limits and errors as well as $s^{*obs}$ come from a single
experiment of finite statistics and as such are subject to statistical
fluctuations.
\section{Comparison with the Bayesian approach}
In the Bayesian approach, an unknown Bayesian prior $P(s)$ is assumed
for the distribution of the parameter $s$ in the absence of any
data. The shape of the prior is guessed at, based on subjective
criteria or using other objective pieces of information.  However,
such a shape is not invariant under transformation of variables. For example,
if we assume that the prior $P(s)$ is flat in $s$, then if we analyze
the problem in $s^2$, we cannot assume it is flat in $s^2$. This
feature of the Bayesian approach has caused controversy. Also, the
notion of a $pdf$ of the data does not exist and $P(c)$ is taken to be a
normalization constant. As such, no goodness of fit criteria exist. 
In the method outlined here, we have used Bayes' theorem to calculate
posterior densities of the fitted parameters while being able to
compute the goodness of fit. The formalism developed here shows that
what is conventionally thought of as a Bayesian prior distribution is
in fact a normalization constant and what Bayesians think of as a
normalization constant is in fact the $pdf$ of the data.
Table~\ref{diff} outlines the major differences between the
Bayesian approach and the new one.
\begin{table}[bht!]
\caption[Differences between the two methods]{
The key points of difference between the Bayesian method and the new method.
\label{diff}}
\centering\leavevmode
\begin{tabular}{|l|l|l|}
\hline
Item & Bayesian Method & New Method \\
\hline
Goodness & Absent & Now available  \\
 of fit  &        & in both binned \\
         &        & and unbinned fits\\
\hline
Data & Used in evaluating  & Used in evaluating \\
     &       theory $pdf$  &  theory $pdf$\\
     &  at data points     &  at data points \\
     &                     & as well as evaluating \\
     &                     &data $pdf$ at data points\\
\hline
Prior & Is a distribution   & No prior needed. \\
      &that is guessed based & One calculates a \\
      &on ``degrees of belief'' & constant from data\\
      &Independent of data, & 
 $\alpha_n = \frac{P^{data}({\bf c_n})}{\int P({\bf c_n}|s) ds}$\\
      &monolithic           & $\rightarrow \infty$ as $n\rightarrow \infty$\\
\hline
Posterior & Depends on Prior. &  Independent of prior. \\
density & & same as frequentists use  \\
$P(s|{\bf c_n})$ & 
$\frac{P({\bf c_n}|s)P(s)}{\int P({\bf c_n}|s)P(s)~ds}$ 
& $\frac{P({\bf c_n}|s)}{\int P({\bf c_n}|s)~ds}$  \\
\hline
\end{tabular}
\end{table}
\section{Further work to be done}
Equation~\ref{pns} can be used to show that the expectation value of $E(s)$
of the parameter $s$ is given by
\begin{eqnarray}
 E(s) = \int s {\cal P}_n(s) ds = \int d{\bf c_n} P({\bf c_n}) \int s P(s|{\bf c_n}) ds\\
=\int  {\bar s}({\bf c_n})P({\bf c_n})d{\bf c_n}
\label{sbar}
\end{eqnarray}
where ${\bar s}({\bf c_n})$ is the average of $s$ for individual
experiments.  Equation~\ref{sbar} states 
$E(s)$ is the weighted average of ${\bar s}({\bf c_n})$ obtained from
individual measurements, the weight for each experiment being the
``data likelihood'' $P({\bf c_n})$ for that experiment.  In the
absence of experimental bias, $E(s)$ would be identical to the true
value $s_T$.  It remains to be shown that the weighted average of
maximum likelihood values $s^*$ from indiviual experiments also
converge to the maximum likelihood point of ${\cal P}_n(s)$.

Also one needs to develop an analytic theory of the goodness of fit for 
unbinned likelihood fits. Finally, one needs to investigate 
a bit more closely  
the transformation properties of ${\cal P}_n(s)$ under change of variable.
\section{Conclusions}

To conclude, we have proposed a scheme for obtaining the goodness of fit 
in unbinned likelihood fits. This scheme involves the usage of two $pdf$'s, 
namely data and theory. In the process of computing the fitted errors, 
we have demonstrated that the quantity in the joint
probability equations that has been interpreted as the ``Bayesian
prior'' is in reality a number and not a distribution. This number is the 
value of the $pdf$ of the parameter, which we call the ``unknown concomitant''
  at the true value of the parameter.
 This number is
calculated from a combination of data and theory and is seen to be an
irrelevant parameter. If this viewpoint is accepted, 
the controversial practice of guessing
distributions for the ``Bayesian Prior'' can now be
abandoned, as can be the terms ``Bayesian'' and ``frequentist''. We show how to use 
the posterior density to rigorously calculate fitted errors.

\begin{acknowledgments}

This work is supported by Department of Energy. The author wishes to thank 
Jim Linnemann and Igor Volobouev for useful comments.

\end{acknowledgments}


\end{document}